\documentclass[12pt]{article}
\begin{document}
\title{Possibility of using dual frequency to control chaotic oscillations of a spherical bubble}
\maketitle
\author{Sohrab Behnia$^1$},
\author{ Amin Jafari$^2$},\author{
Wiria Soltanpoor$^2$} \author{and  Okhtai Jahanbakhsh$^2$}
\begin{center} \noindent
$^1$ Department of Physics, IAU, Urmia, Iran\\

\noindent {$^2$ Department of Physics, Urmia University, Urmia,
Iran}

\noindent Emai: s.behnia@iaurmia.ac.ir
\end{center}
\begin{abstract}
Acoustic cavitation bubbles are known to exhibit highly nonlinear
and unpredictable chaotic dynamics. Their inevitable role in
applications like sonoluminescence, sonochemistry and medical
procedures suggests that their dynamics be controlled. Reducing
chaotic oscillations could be the first step in controlling the
bubble dynamics by increasing the predictability of the bubble
response to an applied acoustic field. One way to achieve this
concept is to perturb the acoustic forcing. Recently, due to the
improvements associated with using dual frequency sources, this
method has been the subject of many studies which have proved its
applicability and advantages. Due to this reason, in this paper, the
oscillations of a spherical bubble driven by a dual frequency
source, were studied and compared to the ones driven by a single
source. Results indicated that using dual frequency had a strong
impact on reducing the chaotic oscillations to regular ones. The
governing parameters influencing its dynamics are the secondary
frequency and its phase difference with the fundamental frequency.
Also using dual frequency forcing may arm us by the possibility of
generating oscillations of desired amplitudes. To our knowledge the
investigation of the ability of using a dual frequency forcing to
control chaotic oscillations are presented for the first time in
this paper.
\end{abstract}
\section{Introduction}
Cavitation bubbles, driven in motion by an acoustic field, are the
example of highly nonlinear forced oscillators. The study of their
behavior, recently, has attracted lots of attention and it is shown
that this phenomenon exhibit highly nonlinear and complex dynamics
both experimentally \cite{Ref1,Ref2,Ref3} and numerically
\cite{Ref4,Ref5,Ref6,Ref7}. Besides their complex behavior,
acoustically driven bubbles have lots of advantageous applications
in material science, sonoluminescence and sonochemistry
\cite{Ref8,Ref9,Ref10,Ref11,Ref12}, sonofusion \cite{Ref13} and
medical procedures such as: lithotripsy, diagnostic imaging, drug
and gene delivery, increasing membrane permeability, opening blood
brain barrier and high intensity focused ultrasound (HIFU) surgery
\cite{Ref14,Ref15,Ref16,Ref17,Ref18}. In all of these applications a
conventional single frequency acoustic source has been used to
provide the bubble oscillations. As an alternative to this
conventional approach recently using dual frequency sources has
lured a great amount of attention in order to improve the outcome of
the applications involving acoustic cavitation. Dual frequency has
been used to enhance sonoluminescence phenomenon
\cite{Ref19,Ref20,Ref21,Ref22,Ref23} resulting in an increase in the
light emission up to $300$ percent. In ultrasound imaging, it
introduces a robust back scatter of ultrasound agents even at low
intensity frequencies, which can be optimized up to $200$ percent
brighter as compared to single frequency employment \cite{Ref24}. In
sonochemistry, using two harmonics has shown to increase the energy
efficacy and enhance sonochemical reactors function
\cite{Ref25,Ref26,Ref27,Ref28}. In therapeutic applications such as
sonodynamic therapy and HIFU, using a dual frequency source is shown
to greatly increase the treatment efficacy
\cite{Ref29,Ref30,Ref31,Ref32}. In the above mentioned applications,
an optimum employment of bubbles suggests that chaotic oscillations
be reduced, because when the bubble motion gets chaotic, its
behavior becomes unpredictable and very hard to control. Therefor,
reducing chaotic oscillations can be the first step in precisely
controlling the bubble dynamics and can provide more beneficial
outcomes. \\In applications involving cavitation, the parameters
like the viscosity, surface tension or the diameter of the bubble
are determined by the media and type of the application. Therefore,
in order to carry out control strategies in the system, the only
remaining parameter that can be dealt with, is the perturbation of
the forcing term. As mentioned using dual frequency sources is
practical and has been proved to have many advantages in an
increasing number of applications. For this reason, in order to
stabilize the bubble motion, we perturb the bubble single frequency
forcing by introducing a secondary frequency drive to the system. In
this respect firstly, through applying the methods of chaos physics,
nonlinear dynamics of a spherical cavitation bubble driven by a
single frequency source was studied in this paper. The considered
control parameters were the applied pressure, frequency and the
bubble initial radius. An evaluation upon the stabilizing effect of
a dual frequency source on chaotic oscillations was carried out. For
a fixed initial frequency, different values of the secondary
frequency along with its phase difference with the fundamental were
studied. These were done through plotting and analyzing the
bifurcation diagrams and the lyapunov exponent spectra. The reason
for using these analysis is that in the absence of any direct and
reliable mathematical methods, bifurcation diagrams and largest
lyapunov exponent spectra are very useful by providing a precise
view on the dynamical behavior of the system over a wide range of
control parameters. They can be used as powerful tools to determine
suitable values for the secondary frequency and the phase
difference. For this reason we analyzed the consequences before and
after applying the introduced method through plotting the
bifurcation diagrams,
lyapunov exponent spectra and time series of the bubble.\\
Results were promising as they indicated an accessible stabilizing
role of a dual frequency source on chaotic oscillations. The
effective control parameters were found to be the value of the
secondary frequency and the phase difference.
\section{The Bubble model} The bubble model used for the numerical
simulation is derived in \cite{Ref4} and is a modification of the
bubble model formulated by Prosperetti\cite{Ref33} from
Keller-Miksis bubble equation\cite{Ref34}. The reason for using this
model is that it generated results in good agreement with those
obtained by solving the full partial differential equations of fluid
dynamics and is suitable for a wide range of amplitudes of
oscillations \cite{Ref35}. It is is given by (1):\\
$$(1-\frac{\dot{R}} {c})R{\ddot{R}}+\frac{3} {2}
{\dot{R}^2}(1-\frac{\dot{R}} {3c})=(\frac{P} \rho)+\frac{1}
{{\rho}c} \frac{d} {dt}(RP)$$
\begin {equation}
 \hspace{5 cm}=(1+\frac{\dot{R}} c)\frac{P}
\rho+\frac{R} {{\rho}c} \frac{dP} {dt}
\end{equation}
with
$$P(R,\dot{R},t)=(P_{stat} -
P_{\nu}+\frac{2\sigma}{R_0})(\frac{R_0}{R})^{3k}- \frac{2\sigma}{R}
- 4{\mu}\frac{\dot{R}} {R}- P_{stat} + P_{\nu} - A$$
\begin{equation}
\end{equation}
where $R_0=10{\mu}m$ is the equilibrium radius of the bubble,
$P_{stat}=100KPa$ is the static ambient pressure, $P_{\nu}=2.33KPa$
is the vapor pressure, $\sigma=0.0725$ $\frac{N}{m}$ is the surface
tension, $\rho=998$ $\frac{Kg}{m^3}$ is the liquid density,
$\mu=0.001$ $\frac{N s}{m^3}$ is the viscosity,
$c$=$1500$$\frac{m}{s}$ is the sound velocity, and $k$=$\frac{4}{3}$
is the polytropic exponent of the gas in the bubble. For the premier
case with single frequency:
\begin{equation}
A=P_a\sin{(2\pi {\nu_1}t)}
\end{equation}
where $\nu_1$ is the frequency of the driving sound field and $P_a$
is the amplitude of the driving pressure.\\
For a dual frequency source:
\begin{equation}
A=P_a[\sin{(2\pi {\nu_1}t)}+\sin{(2\pi {\nu_2}t+\alpha)}]
\end{equation}
where ${\nu_2}$ is the secondary driving frequency and $\alpha$ is
the phase difference.
\section{Results and discussion}
\subsection{Bubble dynamics driven by a single frequency}
The dynamics of the bubble driven by a single frequency source were
studied numerically by solving equation $(1)$. The bifurcation
diagram and the lyapunov exponent spectrum of a bubble with initial
radius of $10 {\mu}m$ were sketched versus pressure and frequency in
a wide range. Figures $1a-b$ show the bifurcation diagram and
lyapunov spectrum of a bubble driven by $500KHz$ of frequency versus
pressure within $10KPa - 2MPa$. Figure $1a$ demonstrates the highly
nonlinear complex dynamics of the bubble throughout the pressure
increase, which includes stable period one, period two, four, eight,
chaos and then period three, period six and once more chaos. Also
there are two windows of complex periodic behavior inside the second
chaotic domain. Figure $1b$ shows its corresponding lyapunov
exponent which confirms figure $1a$ substantiating chaotic behavior
by positive values and stable periodic oscillations by negative ones.\\
Figures $2a-b$ show the bifurcation diagram and lyapunov spectrum of
a bubble when driven by a pressure source of $500KPa$ versus
frequency in the range of $100KHz-2MHz$. We can perceive in figure
$2a$ intermittent occurrence of chaotic and stable behaviors in
bubble oscillations during frequency increase. The transitions from
stable to chaotic oscillations are through period doubling
bifurcations. The last transition to stable everlasting oscillations
occurred through a saddle node bifurcation after an inverse period
doubling. As a quantitative criterion for the comportment
illustrated in figure $2a$, the lyapunov exponent
spectrum was sketched in figure $2b$.\\
The significant undesirable behavior outlined in figures $1$ and $2$
is oscillations which are chaotic. To avoid these unwanted dynamical
effects, it is necessary to carry out control strategies in the
system. To achieve this objective we propose a method called "Dual
frequency technique". In the stated method we introduced a secondary
frequency employment along with the primary driving frequency
source.\\ In this technique, the effect of not only the secondary
frequency but also its phase difference with the basic frequency
must be concomitantly taken into consideration to get the desirable
outcome. More explanation is going to be put forward in this paper.
In order to streamline the manifestation of the method efficacy on
chaotic behaviors, some chaotic zones have been arbitrarily chosen
to be exposed to the secondary frequency. For the associated zones
the dynamical behavior of the bubble was analyzed before and after
control. This is done through analyzing its bifurcation diagram and
the corresponding lyapunov spectrum. Also time series of the
normalized radius of the bubble are presented in order to reveal the
stabilizing effect on the oscillations of the bubble in certain
values of the control parameter.\\
\subsection{Possibility of controlling chaotic oscillations using a dual frequency forcing }
Starting with a completely chaotic zone for a single cavitation
bubble when the control parameter is pressure, figure $3$ has been
presented. The fundamental frequency is $200KHz$ for a bubble of
initial radius of $10{\mu}m$. Figure $3a$ displays the status prior
to applying the technique; while figure $3b$ exhibits the same
system when a secondary frequency of $500KHz$ was applied with the
phase difference of $pi/2$ with the primer one. It is considerable
in figure $3b$ that after applying the secondary frequency, no
chaotic behavior occurred within $1.55 - 2 MPa$, which used to
appear in the system driven by a single frequency. The according
lyapunov spectra have been outlined in figure $4$. The dashed line
corresponds to the situation in the absence of the secondary
frequency. The solid line corresponds to the secondary frequency
employed situation. This figure indicates a significant abatement of
the lyapunov exponent from positive values to negative ones after
the proposed technique was engaged. The controlling phenomenon has
also been granted by plotting the normalized bubble oscillations
versus time in a certain
value of the pressure before and after control in figure $5$.\\
In order to show the capability of the proposed method to control
the chaotic oscillation versus other parameters, we have plotted the
bifurcation diagram of the bubble versus its initial radius, before
and after control, in figures $6a-b$. The frequency and the
amplitude of the driving force are $300KHz$ and $1MPa$,
respectively. A secondary frequency of $1MHz$ is introduced while
the phase difference is $pi/4$. Results indicate its controlling
effect by reducing the chaotic oscillations to regular behaviors of
period $4$ $\longrightarrow$ period $3$ $\longrightarrow$ period $8$
through bubbling bifurcation and again, period $4$ in the initial
radius of the range $37.3{\mu}m$-$39{\mu}m$. Also the normalized
oscillations of a bubble with $37.6{\mu}m$, before and after
applying the secondary frequency, are shown in figures $7a-b$.\\
\subsection{Effects of the phase difference on the dynamics of a dual frequency driven bubble}
In bubble cavitation applications, the optimization of the control
parameters like viscosity and surface tension is mainly imposed by
the media. Other control parameters like pressure, initial frequency
and initial radius are determined by the sort of application. In
dual frequency method the secondary frequency and the phase
difference can be conveniently dealt with to reach a proper choice.\\
Two different circumstances were considered to understand their role:\\
1. Under a constant secondary frequency, its phase difference with
the primary frequency is varied.\\ 2. The amplitude of the secondary
frequency is varied while the phase difference is held constant.\\
For the purpose of studying the effect of phase difference the
chaotic dynamics of a bubble with $4{\mu}m$ of initial radius and
$1MHz$ fundamental frequency, were studied in the range of $1-3MPa$
of pressure. Figure $8a$ shows the resulting bifurcation diagram. It
shows that the bubble dynamics is first chaotic, then becomes of
period four regular oscillations and each of them undergoes a period
doubling bifurcation and again chaotic oscillations occur in the
range of $1.6-3MPa$.  A secondary frequency with the value of
$1.5MHz$ was applied concomitant with the fundamental. The phase
differences of $pi/6$, $pi/2$ and $3pi/4$ were applied. The
resulting bifurcation diagrams are presented in figures $8b-8d$
respectively. As seen in figure $8b$, by applying a phase difference
of $pi/6$ a stable region appeared in the range of about $1.6-2 MPa$
which were chaotic before applying the secondary frequency. When we
applied the phase difference of $pi/2$ (figure$8c$) the bubble
exhibited stable period four oscillations in the range of $1-1.6
MPa$. As seen the oscillations are stable in the range of about
$1-1.3 MPa$ of pressure which they were chaotic before applying the
technique. Figure $8d$ illustrates the case of applying the $3pi/4$
of phase difference. When compared with figure $8a$ we see that a
stable region is formed in the range of about $2-2.5 MPa$. This
domain was completely chaotic prior to applying the dual frequency
method. Also it should be noted that the stable region in the
bifurcation diagram of the bubble driven by a single frequency
source may become chaotic after applying the secondary frequency.
This is obvious in figures $8b$ and $8d$.\\
In summary it is seen that phase difference had a distinguishable
impact on the dual frequency bubble dynamics. Different stable
regions appeared corresponding to typical phase differences. This
suggests certain phase values can be applied to control chaotic
regions of interest.
\subsection{Effects of the secondary frequency on the dynamics of a dual frequency driven bubble}
In order to study the impact of the value of the secondary frequency
on the bubble dynamics, the chaotic oscillations of the bubble in
the previous part (figure $8a$) were chosen. The phase difference
was held constant with the value of $pi/3$ while the secondary
frequency was $500KHz$, $800KHz$, $2MHz$ and $3MHz$. The resulting
bifurcation diagrams are presented in figures $9a-d$, respectively.
Comparing with figure $8a$ using a secondary frequency of $500KHz$
(figure $9a$) has stabilized the first chaotic region in the range
of about $1-1.3MPa$ to period four regular behavior. There are two
small other stable regions in about $1.6-1.7MPa$ and $2.5-2.6MPa$.
As seen in figure $9b$, a small domain of regular behavior around
$1.2MPa$ is created when applying the value of $800KHz$ for the
secondary frequency. Also there is another stable region in the
range of about $1.9-2MPa$. Both of these domains were chaotic before
applying the $800KHz$ of secondary frequency. Figure $9c$ shows the
bifurcation diagram when the secondary frequency is $2MHz$. A vast
region of stable oscillations is generated about $1.9-2.9MPa$ of
pressure. The oscillations are of different periods and some
bubbling bifurcations and period doublings are seen. There are other
stable domains created in the range of about $1.1-1.3MPa$ and
between $1.3-1.4MPa$ of pressure. Employing the secondary frequency
of $3MHz$ (figure $9d$), extremely stabilizes the chaotic
oscillations in the range of approximately $1.8-2.3MPa$ to period
one oscillations. Also another stable region is generated about
$2.6MPa$ of pressure.\\In summary we see that like the influence of
phase difference, the value of the secondary frequency has a strong
impact on the dynamics of a dual frequency driven bubbles. Results
show that by applying certain values for the secondary frequency we
may be capable of controlling the chaotic oscillations in different
regions of interest. Combining the suitable choice for the secondary
frequency and the phase difference can help us greatly for this
purpose.
\subsection{Effects of the secondary frequency and phase difference on the oscillation amplitudes}
As it was seen the maximum oscillation amplitudes may change when
applying the dual frequency technique. In order to study the effects
of variations in frequency and phase difference on the oscillations
amplitude, firstly in figure $10a$ we have plotted time series of
the normalized chaotic oscillations of a bubble with initial radius
of $50{\mu}m$ driven by $300KHz$ of frequency and $1MPa$ of
pressure. Then the proposed method is applied and the obtained
results are presented in figure $10b-i$. In the left column the
phase difference is varied for a fixed secondary frequency and in
the right column the frequency is varied for a constant value of
phase difference. Results demonstrate that certain values of phase
difference and secondary frequency may stabilize the behavior
whereas it can renovate chaos as observed in figures $10c$ and
$10h$. In figure $10$ except for $10c$ and $10h$ the oscillations
have become stable. It is also seen that this technique may arm us
by the
possibility to provide oscillations of desired amplitude.\\
\section{Conclusion}
The dynamics of an acoustically driven gas bubble was studied
applying the method of chaos physics. Results indicated its rich
nonlinear dynamics with respect to variations in the control
parameters of the system. A method based on applying a dual
frequency source is proposed. Simulation results demonstrated that
the proposed procedure may be able to achieve the control objective.
This is possible through choosing appropriate values of the
secondary frequency and its phase difference with initial one.

\clearpage \clearpage 
\newpage
\clearpage
\newpage
The figures of this paper are uploaded in a separate compressed
file. To view the figures please download the compressed file.\\
Figure Captions:\\\\
Figure1. Bifurcation diagram and the corresponding Lyapuonov
spectrum of a bubble with $10{\mu}m$ initial radius driven with
$500KHz$ of frequency versus pressure , $1a$- normalized bubble
radius versus pressure, $1b$- Corresponding Lypunov spectrum.
\\\\
Figure2. Bifurcation diagram and the corresponding Lyapuonov
spectrum of a bubble with $10{\mu}m$ initial radius driven with
$500KPa$ of pressure versus frequency, $2a$- normalized bubble
radius versus frequency $2b$- corresponding Lyapunov spectrum.
\\\\
Figure3. Bifurcation diagrams of the normalized bubble radius driven
by $200KHz$ of frequency with the initial radius of $10{\mu}m$
versus pressure: $3a$- Chaotic behavior before applying the proposed
technique, $3b$- after technique engagement with $\nu_2$=$500KHz$
and $\alpha$=$pi/2$.
\\\\
Figure4. Lyapunov spectra before and after applying the proposed
method. The dashed line represents the case before applying the
method while the solid line represents the system after control.
\\\\
Figure5. Time series of normalized bubble radius driven by $200KHz$
of frequency and $1.7MPa$ of pressure: $5a$- Chaotic oscillations,
 $5b$- Regular oscillations after introducing the dual frequency method ($\nu_2$=$500KHz$ and $\alpha$=$pi/2$).
\\\\
Figure6. Bifurcation diagrams of the bubble normalized radius versus
initial radius driven by $300KHz$ of frequency and $1MPa$ of
pressure: $6a$- chaotic behavior before control, $6b$- Periodic
behavior after the technique engagement ($\nu_2$=$1MHz$ and
$\alpha$=$pi/4$).\\\\
Figure7. Time series of the normalized oscillations of the bubble
with initial radius of $37.6{\mu}m$ driven by $300KHz$ of frequency
and $1MP$ of applied pressure: $7a$- without applying the proposed
technique, $7b$- After applying the proposed technique
($\nu_2$=$1MHz$ and
$\alpha$=$pi/4$).\\\\

Figure8. Bifurcation diagrams of the normalized bubble radius driven
with the fundamental frequency of $1MHz$ and $R_0$=$4\mu m$ versus
pressure: $8a$- Driven by a single frequency. $8b-d$-After applying
the secondary frequency of $1.5MHz$ with the phase difference of
$8b$-$pi/6$, $8c$-$pi/2$,
$8d$-$3pi/4$.\\\\

Figure9. Bifurcation diagrams of the normalized bubble radius driven
by dual frequency source with fundamental frequency of $1MHz$ and
phase difference of $pi/3$ and  $R_0$=$4\mu m$ versus pressure when
the secondary frequencies are: $9a$-$500KHz$, $9b$-$800KHz$,
$9c$-$2MHz$ and $9d$-$3MHz$.\\\\
Figure10. Normalized amplitude of oscillations versus time for a
bubble with initial radius of $50\mu m$ driven by the frequency of
$300KHz$ and $1MPa$ of pressure: $10a$- chaotic oscillations before
technique engagement, left column: after applying the secondary
frequency of $\nu_2 = 1.5 MHz$ while the phase difference is
$10b$-$pi/6$, $10c$-$pi/4$, $10d$-$pi/3$, $10e$-$pi$, right column:
after applying the phase difference of $pi/3$ while the secondary
frequency is $10f$- $400KHz$, $10g$- $600KHz$, $10h$- $1MHz$ and
$10i$-$2MHZ$.\\\\
\end{document}